# Mirror Nuclei of $^{17}$O and $^{17}$F in Relativistic and Non- Relativistic Shell Model

M. Mousavi, M. R. Shojaei*

Department of physics, Shahrood University of Technology, P.O. Box 3619995161-316, Shahrood, Iran

*E-mail: shojaei.ph@gmail.com

**Abstract:** We have investigated energy levels mirror nuclei of the $^{17}$O and $^{17}$F in relativistic and non-relativistic shell model. The nuclei $^{17}$O and $^{17}$F can be modeled as a doubly-magic $^{17}$O=n+(N=Z=8) and $^{17}$F=p+(N=Z=8), with one additional nucleon (valence) in the ld$_{5/2}$ level. Then we have selected the quadratic Hellmann potential for interaction between core and single nucleon. Using Parametric Nikiforov-Uvarov method, we have calculated the energy levels and wave function in Dirac and Schrodinger equations for relativistic and non-relativistic, respectively. Finally, we have computed the binding and excited energy levels for mirror nuclei of $^{17}$O and $^{17}$F and compare with other works. Our results were in agreement with experimental values and hence this model could be applied for similar nuclei.

## 1. Introduction

Special interest resides in the study of masses and sizes for a given element along isotopic chains. Experimentally, their determination is increasingly difficult as one approaches the neutron drip-line; as of today, the heaviest element with available data on all existing bound isotopes is oxygen (Z=8) [1]. Theoretically, the link between nuclear properties and inter-nucleon forces can be explored for different values of N/Z and binding regimes, thus critically testing both our knowledge of nuclear forces and many-body theories [2]. The way shell closures and single-particle energies evolve as functions of the number of nucleons is presently one of the greatest challenges to our understanding of the basic features of nuclei. The properties of single particle energies and states with a strong quasi-particle content along an isotopic chain are moreover expected to be strongly influenced by the nuclear spin-orbit force [3].

The best evidence for single-particle behavior is found near magic (also called closed-shell) nuclei, where the number of protons or neutrons in a nucleus fills the last shell before a major or minor shell gap. For example, the nuclei $^{17}$O and $^{17}$F can be modeled as a doubly-magic $^{17}$O=n+(N=Z=8) and $^{17}$F=p+(N=Z=8), with one additional (valence) nucleon in the ld$_{5/2}$ level. The ground-state spin and parity of $^{17}$O and $^{17}$F are J$^\pi$=5/2$^+$, which corresponds to the spin and parity of the level where the valence nucleon resides [4]. The study of relativistic effects is always useful in some quantum mechanical systems. Therefore, The Dirac equation, which describes the motion of a spin-1/2 particle, has been used in solving many problems of nuclear and high-energy physics. The spin and the pseudo-spin symmetries of the Dirac Hamiltonian were discovered many years ago, however, these symmetries have recently been recognized empirically in nuclear and hadronic spectroscopes [5-6-7].

In this work we use relativistic and non- relativistic shell model for calculation of the energy levels for $^{17}$O and $^{17}$F isotope. Since these isotopes have one nucleon out of the core, Dirac and Schrodinger equations are utilized to investigation them in relativistic and non- relativistic shell model, respectively. These isotopes could be considered as a single particle. We apply the modified Hellmann potential [8, 9] between the core and a single particle because these potentials are important nuclear potentials for a description of the interaction between single nucleon and whole nuclei. Now that the N-N potential is selected, the next step is a solution of the Dirac and Schrodinger equation for the nuclei under investigation. We use the Parametric Nikiforov-Uvarov (PNU) method [29, 30] to solve them.

The scheme of paper is as follows: in Section 2 and 3, energy spectrum in relativistic and non-relativistic shell model is presented respectively. Discussion and results are given in section 4.

## 2. Energy Spectrum in Relativistic Shell Model

In the relativistic description, the Dirac equation of a single-nucleon with the mass moving in an attractive scalar potential S(r) and a repulsive vector potential V (r) can be written as [10]:

$$[-i\hbar c \hat{\alpha}.\hat{\nabla} + \hat{\beta}(Mc^2 + S(r))]\Psi_{n_r,k} = [E - V(r)]\Psi_{n_r,k} \tag{1}$$

Where E is the relativistic energy, M is the mass of a single particle and α and β are the 4 ×4 Dirac matrices.

The wave-functions can be classified according to their angular momentum *j* and spin-orbit quantum number k as follows:

$$\Psi_{n_r,k}(r,\theta,\phi) = \frac{1}{r}\begin{bmatrix} F_{n_r,k}(r)Y^l_{jm}(\theta,\phi) \\ iG_{n_r,k}(r)Y^{\tilde{l}}_{jm}(\theta,\phi) \end{bmatrix} \tag{2}$$

Where $F_{n_r,k}(r)$ and $G_{n_r,k}(r)$ are upper and lower components $Y^l_{jm}(\theta,\phi)$ and $Y^{\tilde{l}}_{jm}(\theta,\phi)$ are the spherical harmonic functions. $n_r$ is the radial quantum number and m is the projection of the angular momentum on the z axis.

Under the condition of the spin symmetry, i. e. Δ(r) = 0 and ∑(r)=V(r)+S(r), the upper component Dirac equation could be written as [11]:

$$\left(-\frac{d^2}{dr^2} + \frac{k(k+1)}{r^2} + \frac{1}{\hbar^2 c^2}[Mc^2 + E][Mc^2 - E + \Sigma(r)]\right)F_{n_r,k}(r) = 0 \tag{3}$$

The quadratic Hellmann potential is defined as [12, 13]:

$$V(r) = -\frac{a}{r} + \frac{b}{r^2}e^{-\alpha r} \tag{4}$$

Where the parameters *a* and *b* are real parameters, these are strength parameters, and the parameter α is related to the range of the potential.

Using the transformation $F_{n,k}(r) = rU_{n,k}(r)$ Eq. (3) brings into the form:

$$U''(r) + \frac{1}{r}U'(r) + \left\{\frac{(E^2 - M^2 c^4)}{\hbar^2 c^2} - \frac{(E + Mc^2)}{\hbar^2 c^2}\Sigma(r) - \frac{k(k+1)}{r^2}\right\}U(r) = 0 \tag{5}$$

Equation (5) is exactly solvable only for the case of k = 0,-1. In order to obtain the analytical solutions of Eq. (5), we employ the improved Pekeris approximation [14] that is valid for *α*≤1. The main characteristic of these solutions lies in the substitution of the centrifugal term by an approximation, so that one can obtain an equation, normally hypergeometric, which is solvable [15,11].

$$\frac{1}{r^2} \approx \frac{\alpha^2}{(e^{-\alpha r} - 1)^2} \tag{6}$$

Also this approximation in reverse order could be used.

We can write the Eq. (5) by using improved Pekeris approximation as summarized below:

$$U''_{n,k}(r) + \frac{2}{r} U'_{n,k}(r) + \frac{1}{r^2}\left[-\chi_2 r^2 + \chi_1 r - \chi_0\right] U_{n,k}(r) = 0 \tag{7}$$

Where the parameters $\chi_2$, $\chi_1$ and $\chi_0$ are considered as follows:

$$\begin{aligned}\chi_2 &= -\gamma \\ \chi_1 &= 2\beta[a + 2b\alpha] \quad ; \quad \gamma = \frac{(E^2 - M^2c^4)}{\hbar^2 c^2}, \quad \beta = \frac{(E + Mc^2)}{\hbar^2 c^2} \\ \chi_0 &= k(k+1) + 2b\beta\end{aligned} \tag{8}$$

Applying PNU method, we obtain the energy equation (with referring to reference [16, 17]) as:

$$(2n+1)\sqrt{\chi_2} + 2\sqrt{\chi_2(\chi_0 + \frac{1}{4})} - \chi_1 = 0 \tag{9}$$

With substituting Eq. (8) in Eq. (9) the energy equation is:

$$(2n+1)\sqrt{\frac{(M^2c^4 - E^2)}{\hbar^2 c^2}} + 2\sqrt{\frac{(M^2c^4 - E^2)}{\hbar^2 c^2}\left(k(k+1) + 2b\frac{(E+Mc^2)}{\hbar^2 c^2} + \frac{1}{4}\right)} - 2\frac{(E+Mc^2)}{\hbar^2 c^2}[a + 2b\alpha] = 0 \tag{10}$$

Let us find the corresponding wave functions. In referring to PNU method in reference [18, 19], we can obtain the upper wave function

$$F_{n_r,k}(r) = N\, r^{\left(\sqrt{\chi_0 + \frac{1}{4}} + \frac{1}{2}\right)} \exp(-\sqrt{\chi_2}\, r) L_n^{\left(2\sqrt{\chi_0 + \frac{1}{4}}\right)}\left[(2 + 2\sqrt{\chi_2})r\right] \tag{11}$$

Where N is the normalization constant, on the other hand, the lower component of the Dirac spinor can be calculated from Eq. (11) as:

$$G_{n_r,k}(r) = \frac{\hbar^2 c^2}{E + Mc^2}(\frac{d}{dr} + \frac{k}{r}) F_{n_r,k}(r) \tag{12}$$

And wave function for Dirac equation can be calculated from Eq. (2) as below:

$$\psi_{n_r,k}(r,\theta,\varphi) = N \begin{bmatrix} Y_{jm}^{l}(\theta,\varphi) \\ \frac{i}{[M + E_{n_r,k}]}\left[\frac{d}{dr} + \frac{k}{r}\right] Y_{jm}^{\tilde{l}}(\theta,\varphi) \end{bmatrix} r^{\left(\sqrt{\chi_0 + \frac{1}{4}} - \frac{1}{2}\right)} \tag{13}$$

$$\exp(-\sqrt{\chi_2}\, r) L_n^{\left(2\sqrt{\chi_0 + \frac{1}{4}}\right)}\left[(2 + 2\sqrt{\chi_2})r\right]$$

Where N is the normalization constant.

## 3. Energy Spectrum in Non-Relativistic Shell Model

The radial Schrodinger equation in spherical coordinates is given as [20, 21]:

$$\frac{d^2 R_{n,l}(r)}{dr^2} + \frac{2}{r}\frac{dR_{n,l}(r)}{dr} + \frac{2\mu}{\hbar^2}\left[E_{n,l} - V(r) - \frac{\hbar^2}{2\mu}\left(\frac{\ell(\ell+1)}{r^2}\right)\right] R_{n,l}(r) = 0 \tag{14}$$

By substituting quadratic Hellmann potential in the equation (14) the radial Schrodinger equation is reduced as following:

$$\frac{d^2 R_{n,l}(r)}{dr^2} + \frac{2}{r}\frac{dR_{n,l}(r)}{dr} + \frac{2\mu}{\hbar^2}\left[E_{n,l} + \frac{a}{r} - \frac{b}{r^2}e^{-\alpha r} - \frac{\hbar^2}{2\mu}\left(\frac{\ell(\ell+1)}{r^2}\right)\right]R_{n,l}(r) = 0 \tag{15}$$

Since the Schrodinger equation with above potential has no analytical solution for l≠0 states, an approximation has to be made. Using improved Pekeris approximation [15,11] that is presented in the previous section the Eq. (15) is as following:

$$\frac{d^2 R_{n,l}(r)}{dr^2} + \frac{2}{r}\frac{dR_{n,l}(r)}{dr} + \frac{1}{r^2}\left[-\chi'_2 r^2 + \chi'_1 r - \chi'_0\right]R_{n,l}(r) = 0 \tag{16}$$

Where the parameters $\chi'_2$, $\chi'_1$ and $\chi'_0$ are considered as follows:

$$\chi'_2 = -\frac{2\mu}{\hbar^2}E$$
$$\chi'_1 = \frac{2\mu}{\hbar^2}[a + b\alpha] \tag{17}$$
$$\chi'_0 = \frac{2\mu}{\hbar^2}b + \ell(\ell+1)$$

Applying PNU method, we obtain the energy equation (with referring to the reference [16, 17]) as below:

$$(2n+1)\sqrt{\chi'_2} + 2\sqrt{\chi'_2(\chi'_0 + \frac{1}{4})} - \chi'_1 = 0 \tag{18}$$

Substituting Eq. (17) in Eq. (18) the energy equation is:

$$E_{n,\ell} = -\frac{2\mu}{\hbar^2} \frac{(a+b\alpha)^2}{\left[(2n+1) + 2\sqrt{\frac{2\mu}{\hbar^2}b + \ell(\ell+1) + \frac{1}{4}}\right]^2} \tag{19}$$

And radial wave function for Schrodinger equation can be calculated with referring to PNU method in reference [16, 17] as below:

$$R_{n_r,k}(r) = N'\, r^{\left(\sqrt{\chi'_0 + \frac{1}{4}} - \frac{1}{2}\right)} \exp\left(-\sqrt{\chi'_2}\, r\right) L_n^{\left(2\sqrt{\chi'_0 + \frac{1}{4}}\right)}\left[\left(2 + 2\sqrt{\chi'_2}\right)r\right] \tag{20}$$

Where N′ is the normalization constant.

## 4. Result and Discussion

We consider mirror nuclei $^{17}O$ and $^{17}F$ isotopes with a single nucleon on top of the $^{16}O$ and $^{16}F$ isotopes core. Since these isotopes have one nucleon out of the core, these isotopes could be considered as single particle model in relativistic and non- relativistic shell model. Relativistic mean field (RMF) theory, as a covariant density functional theory, has been successfully applied to the study of nuclear structure properties [22]. In the relativistic mean field theory, repulsive and attractive effects at the same time have been combined, via vector and scalar potentials, also it involves the antiparticle solutions and spin-orbit interaction [23]. So we could use of Dirac equation for investigation them.

The ground state and first excited energies of mirror nuclei $^{17}$O and $^{17}$F isotopes are obtained in relativistic and non-relativistic shell model by using Eq. (10) and Eq. (19), respectively. These results for relativistic and non-relativistic shell model are compared with the experimental data and others work in table 1.

**Table 1**: the ground state and the first excited energy of $^{17}$F and $^{17}$O isotopes in non-relativistic and relativistic (with α=0.012fm$^{-1}$).

| Isotope | state | E-Our(MeV) Non-Relativistic | E-Our(MeV) Relativistic | E-Other(MeV) [24] | E-Exp.(MeV) [25] |
|---|---|---|---|---|---|
| $^{17}$F | 1d$_{5/2}$ | -128.6460 | -128.5116 | 129.14 | -128.2196 |
|  | 2s$_{1/2}$ | -128.2364 | -128.0045 | ------ | -127.7243 |
| $^{17}$O | 1d$_{5/2}$ | -132.1423 | -131.9427 | 132.88 | -131.7624 |
|  | 2s$_{1/2}$ | -131.3213 | -131.0455 | ------ | -130.8916 |

The difference between excited state energies and ground state energies of mirror nuclei $^{17}$O and $^{17}$F isotopes for relativistic and non-relativistic shell model are compared with the experimental data and others work in table 2.

**Table 2.** Comparison of the energies of excited states of $^{17}$O and $^{17}$F, relative to the ground-state energies (the $\left(5/2\right)_1^+$ state of $^{17}$O and $^{17}$F), all entries are in MeV.

| Isotope | Excited state | Others work N$^3$LO [26] | Others work CD-Bonn [27] | Others work V$_{18}$ [28] | Our work Non-Relativistic | Our work Relativistic | Exp. [25] |
|---|---|---|---|---|---|---|---|
| $^{17}$F | $\left(1/2\right)_1^+$ | 0.428 | 0.805 | 0.062 | 0.4096 | 0.5071 | 0.495 |
| $^{17}$O | $\left(1/2\right)_1^+$ | -0.025 | 0.311 | -0.390 | 0.8210 | 0.8972 | 0.870 |

The calculated energy levels have good agreement with experimental values. Therefore, the proposed model can well be used to investigate other similar isotopes and compare with experimental data.

The authors declare that there is no conflict of interest regarding the publication of this paper.